\documentclass[prl,twocolumn,draft,amsmath,showpacs]{revtex4}
\usepackage{graphics}

\begin{document}
%%%%%%%%%%%%%%%%%%%%%%%%%%%%%%%%%%%%%%%%%%

\bibliographystyle{prsty}
\input epsf

\title {Magneto-optical reflectivity of superconducting $MgB_2$
single crystals}

\author {A. Perucchi, L. Degiorgi, J. Jun, M. Angst and J. Karpinski}
\affiliation{Laboratorium f\"ur Festk\"orperphysik, ETH Z\"urich,
CH-8093 Z\"urich, Switzerland}

\date{\today}

\begin{abstract}
We present magneto-optical reflectivity results in the basal-plane of
the hexagonal $MgB_2$. The data were collected on a mosaic of $MgB_2$
single crystals with $T_{c}=38 ~K$ from the ultraviolet down to the far
infrared as a function of temperature and magnetic field oriented
along the c-axis. In the far infrared there is a clear signature of
the superconducting gap with a gap-ratio $2\Delta/k_{B}T_{c}\sim
1.2$, well below the
weak-coupling value. The gap is suppressed in an external magnetic
field, which is a function of temperature. We extract the upper
critical field $H_{c2}$ along the c-axis. The temperature dependence of
$H_{c2}$ is compatible with the Helfand-Werthamer behaviour.
\end{abstract}
\maketitle

Since the discovery of superconductivity in $MgB_{2}$ with T$_c\sim 40
~K$ \cite{ref1},
substantial experimental activity has been devoted to elucidate the
nature of its superconducting state. The issue is to assess whether
$MgB_{2}$ is one of the best-optimized Bardeen-Cooper-Schrieffer (BCS)
materials or whether its superconducting properties stem from a novel
pairing mechanism. The amplitude of the superconducting gap and its
spectroscopic signature, as a function of temperature and magnetic
field, are expected to depend on the driving mechanism \cite{ref2}.

Infrared spectroscopy is a very promising and powerful method for the
observation of the gap. Optical experiments, besides being
contact-less techniques, offer the important advantage that the
electromagnetic radiation penetrates deeply inside the bulk. Analysis
of previous optical data \cite{ref3,ref4,ref5,ref6} revealed gap
values in the range
$2\Delta \sim 3-6  ~meV$, which are rather small compared to the BCS estimate
$2\Delta
=12  ~meV$ \cite{ref2}. The range of superconducting gaps proposed for
$MgB_{2}$ is
even wider when considering other techniques, such as Raman or
tunneling spectroscopy, photoemission or specific heat
\cite{ref7,ref8,ref9,ref10,ref11}, which
yield values of $2\Delta$ between $3$ and $10 ~meV$. This wide spread of gap
values is also accompanied in some cases by deviations from the BCS
temperature dependence, pointing to complex behaviour in this novel
superconductor. Possible origins of the large gap distribution might
include effects associated with the sample quality (i.e.,
imperfections and distribution of the surface composition) and with
the specimen type (i.e., polycrystalline or films). More intrinsic
origins of the wide range of gaps comprise the existence of an
anisotropic (s-wave) superconducting gap as well as a multi-gap
mechanism \cite{ref12}. It has been also conjectured that the different
values of the gap might be partly explained by the preferential
sensitivity of particular experiments to different portions of the
gap distribution \cite{ref12}.

Here, we present infrared reflectivity data $R(\omega)$ on high quality
\emph{single} crystals of $MgB_{2}$ with well-characterized optical surface
parallel to the basal-plane. We focus our attention to the dependence
of $R(\omega)$ to the magnetic field oriented along the c-axis at selected
temperatures close to and below $T_{c}$. The corresponding real part of
the optical conductivity exhibits a depletion of oscillator strength
due to the opening of the superconducting gap. The gap size is less
than half the value expected from an isotropic BCS model. By applying
magnetic fields along the c-axis we induce a suppression of
superconductivity, as suggested by the disappearance of the gap. We
extract the upper critical field $H_{c2}$ along the c-axis and find it to
be compatible with the Helfand-Werthamer \cite{ref13} calculation. This seems
to indicate that no unconventional mechanism needs to be invoked to
explain the bulk upper critical field. Moreover, our magneto-optical
results suggest a dirty-limit metallic normal state.

Single crystals of $MgB_{2}$ with $T_{c}= 38  ~K$ were grown using a
high-pressure cubic anvil cell and were synthesized from a precursor
mixture of $99.9\%$ pure Mg powder and $97\%$ pure amorphous boron. The
crystals were grown in a BN container of $6 ~mm$ diameter under a
pressure of $30  ~kbar$. The temperature was increased up to $1800 ~^0 C$ and
kept constant for $0.5  ~h$. The temperature was then decreased and
pressure released. Flat crystals of $MgB_{2}$ of a size up to
1.2x0.8x0.05$ ~mm^3$ have been obtained \cite{ref14}. We built a
mosaic consisting
of three crystallites from the same batch and with optical surface
oriented perpendicular to the c-axis.

We measured the optical reflectivity $R(\omega)$ in the basal-plane of the
hexagonal $MgB_{2}$. The mosaic had an optical measurable surface of
2x2$ ~mm^2$. The largest crystallite has been placed in the center of the
mosaic. A series of measurements identifying different spots on the
specimen were performed and equivalent results were collected,
excluding important effects due to the surface scattering. In the far
infrared (FIR) spectral range, data were collected as a function of
temperature ($1.5-300  ~K$) and with varying magnetic field H ($0-7  ~T$)
oriented along the c-axis with possible misalignment of less than $5
^{0}$.
The FIR spectral range (i.e., $20-600  ~cm^{-1}$) was covered with a Bruker
Fourier spectrometer equipped with He-cooled Ge bolometer detector
and with an Oxford magnet cryostat with appropriate optical windows.
We present our FIR results down to approximately $25  ~cm^{-1}$, the lower
frequency limit where the collected data are reproducible over
several runs and in different combinations of temperature and
magnetic field. Because of the mosaic configuration of our specimen
as well as the residual surface scattering and diffraction limit
effects, we cannot fully trust the data below $25
~cm^{-1}$.

%<<<<<<<<<<<<<<<<<<<<<<<< FIGURE 1 >>>>>>>>>>>>>>>>>>>>>>>>>
Figure 1 provides an overview, showing the magnetic field dependence
of $R(\omega)$ in FIR at $1.6$ and $20  ~K$ [15]. The normal state
$R(\omega)$ at $40 ~K$
and $0 ~T$ is shown in Fig. 1, as well. At zero field $R(\omega)$ at
$T<T_{c}$ is
remarkably enhanced below $50  ~cm^{-1}$: $R(\omega)$ at $1.6  ~K$
reaches the total
reflection at about $30  ~cm^{-1}$, while $R(\omega)$ at $20  ~K$ is
still enhanced but
only approaches the total reflection. By applying a magnetic field
along the c-axis, the obvious trend at all temperatures is to reduce
the enhancement of $R(\omega)$ (e.g., already at $1.6  ~K$ and $4
~T$ $R(\omega)$ clearly
deviates from the total reflection behaviour seen at $0  ~T$). At any
temperature, there is a specific magnetic field for which $R(\omega)$
recovers the normal state shape. We have also verified the field
independence of $R(\omega)$ above $38  ~K$. It is worth mentioning that our FIR
results in zero field merge nicely with the submillimeter range data
of Pimenov {\it et~al.} obtained with a $MgB_{2}$ film \cite{ref3}.
As it will be
stressed below, the change of slope in $R(\omega)$ for $T<T_{c}$ with
respect to
the normal state behaviour (i.e., at $40 ~K$) is associated to the
optical manifestation of the superconducting gap (arrow in Fig. 1a).
While the overall $R(\omega)$ shape below $T_{c}$ is compatible with
the BCS-like
behaviour \cite{ref2}, we note the rather pronounced drop of $R(\omega)$ in
FIR.
This leads to a shallow minimum in $R(\omega)$ at about $100  ~cm^{-1}$
\cite{ref15}, which
bears a striking similarity with the result on thin film \cite{ref3}. In
passing, the consequence of such a minimum is an excitation spectrum,
which may be interpreted within a multiband (i.e., two-Drude
component) scenario \cite{ref15,ref16,ref17}. Above $50  ~cm^{-1}$
the spectra were found to
be temperature and magnetic field independent. The inset of Fig. 1
displays an overall view of $R(\omega)$ at $300 ~K$, measured up to
$10^5  ~cm^{-1}$,
which is in fair agreement with $R(\omega)$ on films and dense
polycrystalline samples \cite{ref3,ref16,ref18}. Therefore, the $300
~K$ reflectivity
was used in the Kramers-Kronig transformation in order to calculate
the optical conductivity. To this end, the standard extrapolation was
employed above $10^5  ~cm^{-1}$ \cite{ref19}. Below $25  ~cm^{-1}$ $R(\omega)$
at $T>T_{c}$ was
extrapolated using the Hagen-Rubens law \cite{ref19}. Below $T_{c}$,
$R(\omega)$ was
approximated to $100\%$ when $R(\omega)$ reached total reflection at finite
frequency in the measured spectral range, otherwise $R(\omega)$ was assumed
to gradually increase to $100\%$ for $\omega\to 0$. Alternative extrapolation
methods were checked for $\omega\to 0$ but this issue does not affect the main
argument of our discussion.

Figure 2 shows the magnetic field dependence of the ratio of the real
part $\sigma_{1}(\omega)$ of the optical conductivity to its normal
state value (at
$40  ~K$) at selected temperatures. Focusing first on the $0  ~T$ data, we
observe the strong depletion of $\sigma_{1}(\omega)$ for $T<T_{c}$.
The conductivity is
very small or even zero (i.e., as a consequence of the total
reflection for $T\ll T_{c}$ (Fig. 1)) below $30  ~cm^{-1}$ and then increases
monotonically at higher photon energies. We identify the frequency,
where the onset of absorption (i.e., increase of $\sigma_{1}(\omega)$,
Fig. 2) occurs, as
the superconducting gap value. The superconducting gap $2\Delta\sim 31
  ~cm^{-1}$
($3.8  ~meV$) at $T\ll T_{c}$ for our $MgB_{2}$ single crystal agrees
fairly well
with those previously achieved with optical methods on films ($3-5
  ~meV$
\cite{ref3,ref4,ref5}), as well as on polycrystalline samples ($3-4  ~meV$
\cite{ref6}). The
resulting gap-ratio $2\Delta/k_{B}T_{c}\sim 1.2$ seems to confirm the
trend of an
optical superconducting gap in $MgB_{2}$ substantially below the estimate
of the weak-coupling BCS theory \cite{ref2}.

%<<<<<<<<<<<<<<<<<<<<<<<< FIGURE 2 >>>>>>>>>>>>>>>>>>>>>>>>>
There is an increasing consensus for a double gap scenario
\cite{ref20,ref21};
the larger gap with $2\Delta/k_{B}T_{c}\sim 4$ associated with the
two-dimensional
$\sigma$-band and the smaller one with $2\Delta/k_{B}T_{c}\sim 1.3$
associated with the
three-dimensional $\pi$-band \cite{ref22}. All optical data consistently show a
single gap feature \cite{ref3,ref4,ref5,ref6} and seem to rather
yield the lower limit of
the gap distribution. While theoretical calculations of the optical
conductivity within a multi-gap scenario would be imperative, we note
another intriguing aspect of our data: the optical conductivity rises
in a s-wave like manner but much more steeply than a Mattis-Bardeen
(MB) calculation for photon energies larger than the gap
\cite{ref2,ref23}. A
similar situation was encountered already in the superconducting
fullerenes \cite{ref24}. It was shown that the Eliashberg formalism can
explain this behaviour. It goes beyond the MB calculation by taking
into account a realistic phonon spectrum (i.e., also phonon assisted
absorptions, like in the Holstein process) and arbitrary impurity
scattering \cite{ref15,ref24}.

Another key aspect of this study is the possibility to exploit bulk
sensitive and contact-less optical data in order to extract the upper
critical field $H_{c2}$ along the c-axis. The fields for which the normal
state is recovered (i.e., the conductivity in Fig. 2 yield a ratio
close to one) represent the optical counterpart of $H_{c2}$ along the
c-axis. These characteristic fields are shown in Fig. 3 \cite{ref25}. While
the temperature-grid considered here does not allow any conclusion
about possible anomalous curvature of $H_{c2}(T)$ (particularly close to
$T_{c}$), it is safe to say that our values of $H_{c2}$ are compatible with the
temperature dependence derived by Helfand and Werthamer (HW) (Fig. 3)
\cite{ref13}. This also agrees with the trend seen in other experiments
\cite{ref26,ref27,ref28}. The roughly linear temperature dependence
of $H_{c2}$ above $20  ~K$,
besides being in broad agreement with the BCS theory \cite{ref2}, allows the
determination of $H_{c2}(0)$ within the HW approach:
$H_{c2}(0)\sim 0.7(dH_{c2}/dT)T_{c}=5.2  ~T$ \cite{ref13}. This value
is in accord with the
$T\to 0$ limit of the optical estimation of $H_{c2}$. Our values of
$H_{c2}$ along
the c-axis are somehow larger than those estimated from the thermal
conductivity \cite{ref26}, torque magnetometry \cite{ref27}, magnetization and
specific heat \cite{ref28} on single crystals. As far as the comparison with
magneto-transport data is concerned, our $H_{c2}$ values agree perfectly
with the most recent data of Welp {\it et~al.} \cite{ref28} but are
substantially
lower than those of Sologubenko {\it et~al.} \cite{ref26}. In order
to explain the
discrepancy between $H_{c2}$ from the thermal conductivity and from the
resistivity, Sologubenko {\it et~al.} argued that surface effects are
important in relation to superconductivity.

%<<<<<<<<<<<<<<<<<<<<<<<< FIGURE 3 >>>>>>>>>>>>>>>>>>>>>>>>>
 From $H_{c2}(0)\sim 5.2  ~T$ along the c-axis, we can calculate the coherence
length $\xi$ in the basal-plane \cite{ref13}:
$\xi(0)=(0.7\Phi_{0}/2\pi H_{c2}(0))^{1/2}\sim 7 ~ nm$, a
value in broad agreement with estimations from other experiments
\cite{ref12,ref26,ref28}. For comparison, the BCS coherence length,
calculated with
the expression $\xi_{0}=\hbar v_{F}/\pi\Delta (0)\sim 52  ~nm$ using
the optical
gap $\Delta(0)\sim 15 ~cm^{-1}$,
is three times larger than the value $\xi_{0}=\hbar v_{F}/\pi
1.76k_{B}T_{c}\sim  17 ~nm$ for T$_{c}=38
~K$ and the Fermi velocity $v_{F}=$4.8x$10^7  ~cm/sec$ [2,22]. Obviously, these
values of $\xi_{0}$ suffer from the fact that the gap ratio is sizeably
smaller than the weak-coupling limit prediction (i.e.,
$2\Delta=3.52k_{B}T_{c}$).
In accordance with previous optical investigation \cite{ref4,ref5,ref16}, our
normal state properties suggests a relaxation time $\tau\sim$
1.8x$10^{-14} ~sec$
\cite{ref15}, which leads to the mean free path $l=v_{F}\tau =8.5  ~nm$.
Our value of $l$
is smaller than or at most comparable to the coherence length. This,
together with the fact that the optical scattering rate $\Gamma\sim
1/\tau\ge 2\Delta$,
puts $MgB_{2}$ among the so-called dirty limit superconductors
\cite{ref2}. All
optical experiments so far \cite{ref3,ref4,ref5,ref6,ref16,ref18}
unanimously support the dirty
limit scenario. However, this is a controversial and puzzling issue,
since magnetic and thermodynamic probes are more in favor of the
clean limit \cite{ref26,ref28}.

In summary, we have performed magneto-optical measurements on
$MgB_{2}$
single crystals. Besides identifying the signature of the
superconducting gap with $2\Delta/k_{B}T_{c}\sim 1.2$, we have also
extracted $H_{c2}$
along the c-axis. It remains to be seen how one can reconcile an
overall BCS-like $H_{c2}$ with less conventional manifestations of the
superconducting state, like the small gap value. Moreover, the
controversy about the clean versus dirty limit scenario awaits
resolution.\\

\acknowledgments
We wish to thank A. Sologubenko, R. Monnier, I. Landau, S. Broderick
and H.R. Ott for fruitful discussions. This work has been financially
supported by the Swiss National Foundation for the Scientific
Research.

\newpage

%<<<<<<<<<<<<<<<<<<<<<<<< FIGURE 1 >>>>>>>>>>>>>>>>>>>>>>>>>
\begin{figure}[t]
   \begin{center}
    \leavevmode
    \epsfxsize=1\columnwidth \epsfbox {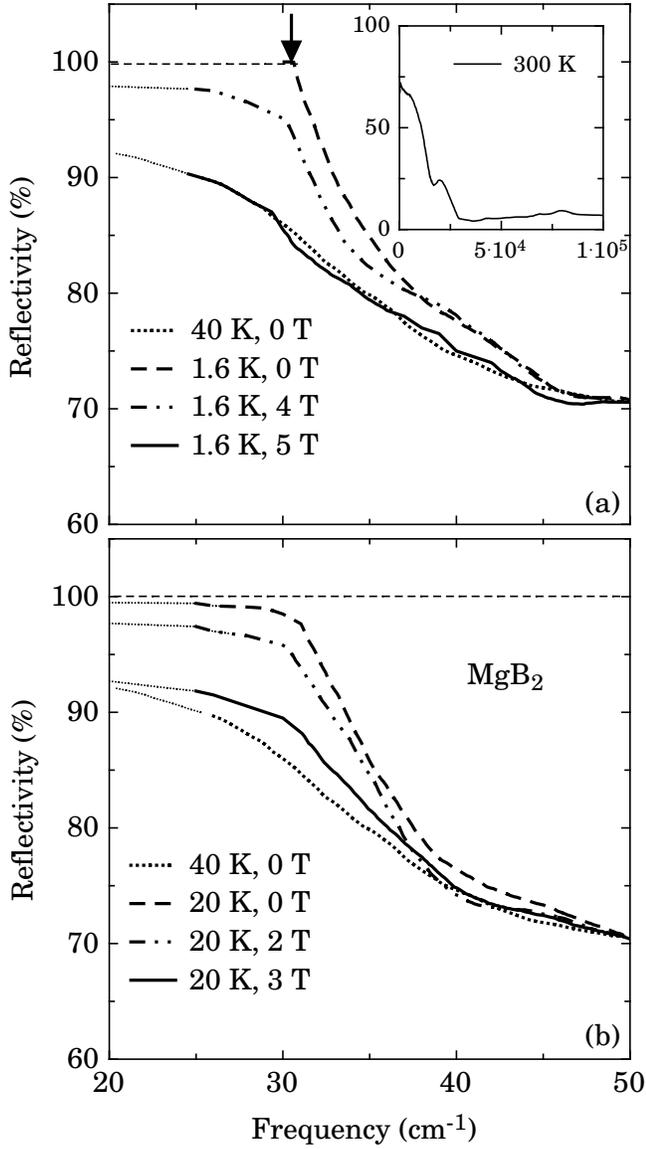}
     \caption{
     Magnetic field dependence $(H\parallel c)$ of the far infrared optical
     reflectivity $R(\omega)$,
    measured in the basal-plane, of a mosaic of $MgB_{2}$ single crystals
    at (a) $1.6  ~K$ and
    (b) $20  ~K$. The reflectivity at $40  ~K$ in zero field is
shown in both figures, as
    reference for the normal state behaviour. The arrow marks the
energy where total
    reflection is reached at $1.6  ~K$, as signature for the
superconducting gap. The thin
    dotted line highlights the extrapolation towards zero energy
for the Kramers-
    Kronig transformation. The inset in part (a) gives the
overall behaviour
    of $R(\omega)$ at
    $300  ~K$ up to $10^5  ~cm^{-1}$.
}
\label{R}
\end{center}
\end{figure}
%<<<<<<<<<<<<<<<<<<<<<<<< figure 1 >>>>>>>>>>>>>>>>>>>>>>>>>

%<<<<<<<<<<<<<<<<<<<<<<<< FIGURE 2 >>>>>>>>>>>>>>>>>>>>>>>>>
\begin{figure}[t]
   \begin{center}
    \leavevmode
    \epsfxsize=1\columnwidth \epsfbox {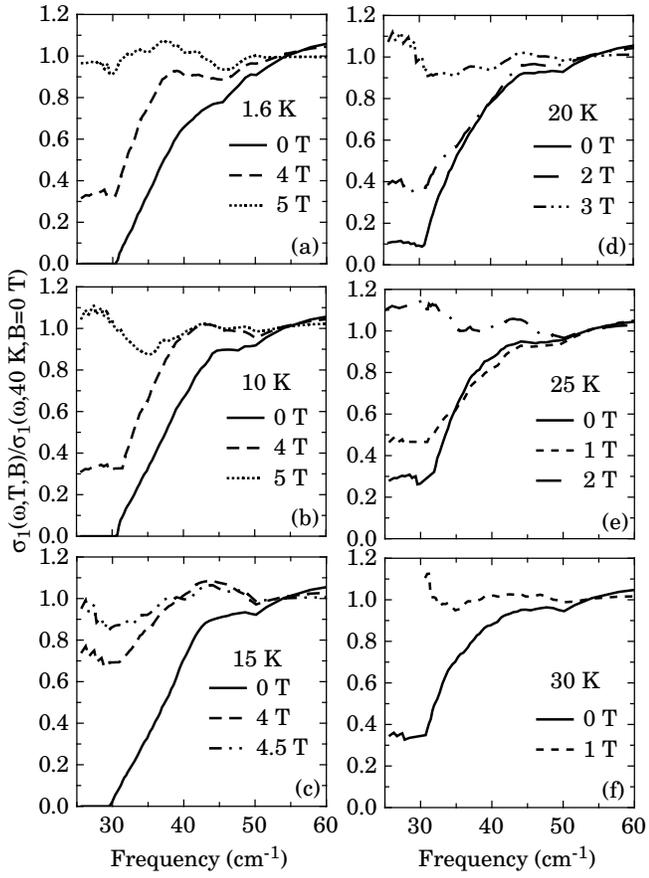}
     \caption{
      Magnetic field dependence ($H\parallel c$) of the conductivity ratio
      $\sigma_{1s}/\sigma_{1n}$ at selected
    temperatures in the FIR spectral range. The normal state
    $\sigma_{1n}$ corresponds to the
    measurement at $40  ~K$ in zero field.
}
\label{sigma}
\end{center}
\end{figure}
%<<<<<<<<<<<<<<<<<<<<<<<< figure 2 >>>>>>>>>>>>>>>>>>>>>>>>>

%<<<<<<<<<<<<<<<<<<<<<<<< FIGURE 3 >>>>>>>>>>>>>>>>>>>>>>>>>
\begin{figure}[t]
   \begin{center}
    \leavevmode
    \epsfxsize=1\columnwidth \epsfbox {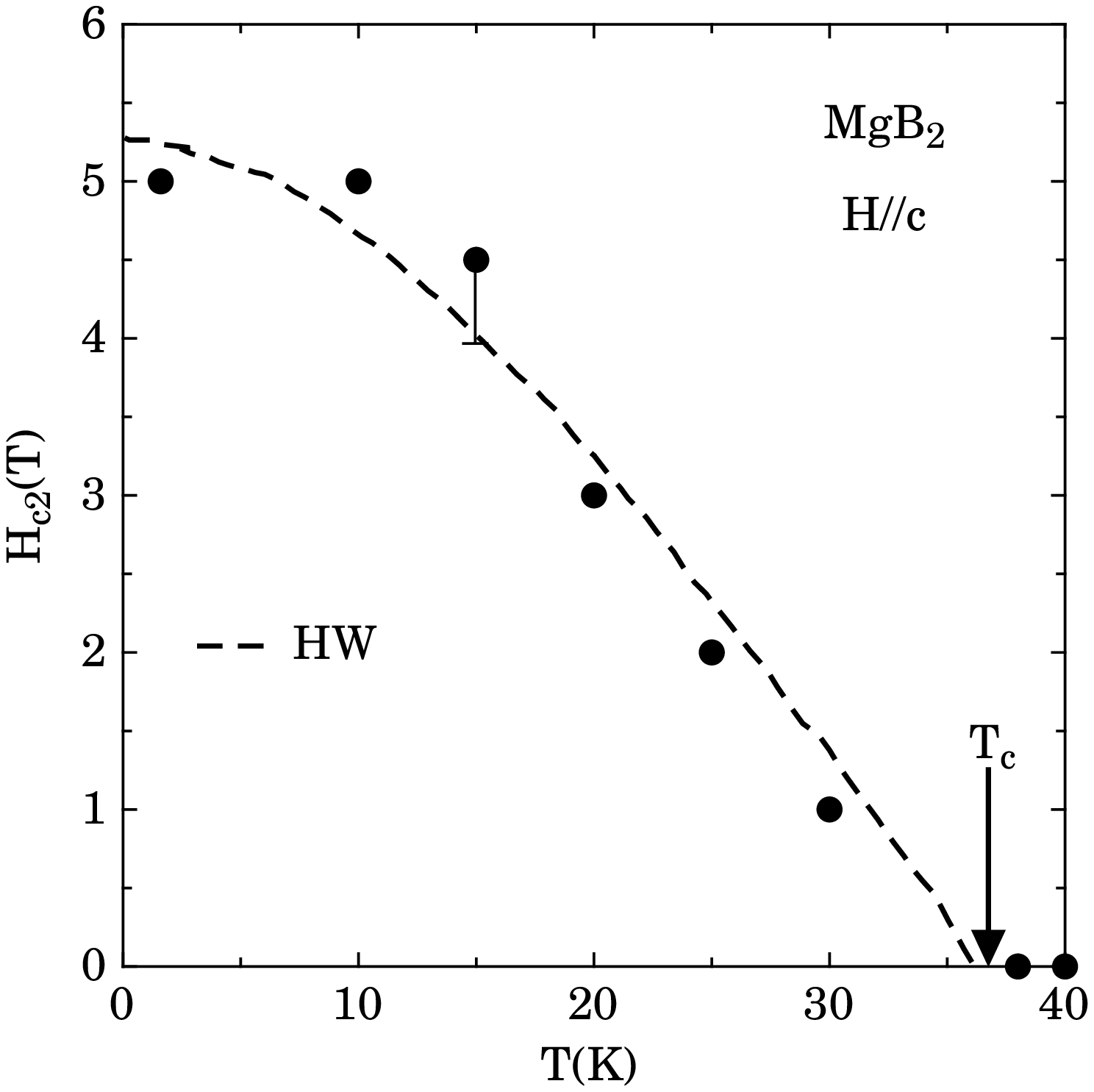}
     \caption{
   Temperature dependence of $H_{c2}$ along the c-axis, evaluated from
the optical
    experiment. The thick dashed line is compatible with
calculations due to Helfand
    and Werthamer (HW) \cite{ref13}.
}
\label{Hc2}
\end{center}
\end{figure}
%<<<<<<<<<<<<<<<<<<<<<<<< figure 3 >>>>>>>>>>>>>>>>>>>>>>>>>


\begin{thebibliography}{10}
\bibitem{ref25}
We note that the point at $15 ~K$ corresponds to the upper value of
$H_{c2}$ evaluated from the optical spectra. For a few measurements the
normal state at $15 ~K$ was
already reached at $4 ~T$. This is shown by the error bar in Fig. 3.
For the other
temperatures the estimation of the $H_{c2}$ fields was unique.
\end{thebibliography}
\end{document}